\documentclass[aps,prb,reprint,superscriptaddress,floatfix]{revtex4-1}

\usepackage{amssymb,amsmath,bm,graphicx}
\usepackage{ifpdf,color,mathrsfs}

\definecolor{darkblue}{rgb}{0,0,.6}
\definecolor{darkgreen}{rgb}{0,0.5,0}

\ifpdf
  \usepackage{epstopdf}
  \usepackage[pdftex,unicode,
  pdfstartview={FitH},pdfborder={0 0 0}]{hyperref}
  \usepackage{hypcap}
\else
  \usepackage[hypertex]{hyperref}
\fi
\hypersetup{
  bookmarksnumbered = true,
  colorlinks = true, linkcolor = darkblue,
  citecolor = darkblue, filecolor = darkblue,
  menucolor = darkblue, urlcolor = darkblue
}

\newcommand{\iu}{i} 
\newcommand{\de}{d} 
\newcommand{\ee}{e} 

\let\Re\undefined
\let\Im\undefined
\DeclareMathOperator{\Re}{Re}
\DeclareMathOperator{\Im}{Im}

\begin{document}

\title{Ultrafast optical Faraday effect in transparent solids}

\author{Michael S.~Wismer}
\email[]{michael.wismer@mpq.mpg.de}
\affiliation{Max-Planck-Institut f\"ur Quantenoptik, Hans-Kopfermann-Str. 1, 85748 Garching, Germany}

\author{Mark I.~Stockman}
\email[]{mstockman@gsu.edu}
\affiliation{Center for Nano-Optics (CeNO) and Department of Physics and Astronomy, Georgia State University, Atlanta, Georgia 30340, USA}

\author{Vladislav S.~Yakovlev}
\email[]{vladislav.yakovlev@mpq.mpg.de}
\affiliation{Max-Planck-Institut f\"ur Quantenoptik, Hans-Kopfermann-Str. 1, 85748 Garching, Germany}
\affiliation{Ludwig-Maximilians-Universit\"at, Am Coulombwall~1, 85748 Garching, Germany}

\date{\today}

\begin{abstract}
	We predict a strong-field ultrafast optical Faraday effect, where a circularly polarized ultrashort optical pulse induces transient chirality in an achiral transparent dielectric.
	This effect is attractive for time-resolved measurements because it gives access to the non-instantaneity of the nonlinear medium response, and also because it represents relaxation of time-reversal symmetry by all-optical means.
	We propose probing the induced transient chirality with a weak linearly polarized ultraviolet pulse that is shorter than the near-infrared pump pulse.
	The predicted effects are ultrafast: the induced chirality vanishes for probe delays exceeding the duration of the near-infrared pulse.
	This opens up possibilities for applications in ultrafast circular-polarization modulators and analyzers.
\end{abstract}


\maketitle

\section{Introduction}
In an optically inactive (achiral) isotropic medium, a weak linearly polarized light pulse preserves its polarization state as it propagates.
However, even in such media, nonlinear interaction with a strong circularly polarized laser beam may rotate the polarization plane of the probe pulse, which is known as the \emph{optical Faraday effect}~\cite{Atkins_MP_1968, Perlin_FTT_1980, Perlin_JETP_1994}.
Similarly to the Faraday effect and in contrast to optical activity in chiral media, the rotation angle, $\Delta\theta$, changes its sign if the propagation direction of the probe pulse is reversed.
That is, the circularly polarized strong field relaxes the time-reversal ($\mathcal{T}$) symmetry.
If such a field also induces nonlinear absorption, it may change the ellipticity of the propagated probe pulse.
This class of phenomena was first discovered in atomic vapors~\cite{Arutiunian_ZhETF_1975,Liao_PRA_1977,Thirunamachandran_CPL_1977}.
In those measurements, the frequencies of pump and probe pulses were tuned to atomic transitions, which enhanced the nonlinear interaction and, at the same time, rendered it non-parametric.
Light-induced ellipticity and polarization rotation were investigated for solids in the parametric and non-parametric cases, where the medium was transparent to either both laser pulses~\cite{Danishevskii_JETF_1981, Perlin_JETP_1994} or just the pump pulse~\cite{Popov_OL_1994,Svirko_1998_Polarization_book}.
To the best of our knowledge, the optical Faraday effect has never been investigated with femtosecond pulses.
Searching for new approaches to ultrafast manipulation of light with light, we question how the transfer of angular momentum from a pump pulse to a transparent achiral solid and then to a probe pulse occurs on an attosecond time scale.

The nonlinear effects that we study become particularly significant for intense few-cycle laser pulses.
Such pulses enable nondestructive measurements \cite{Schultze_et_al_Nature_2012_Controlling_Dielectrics} at peak intensities up to $\sim 10^{14}\ \mbox{W}/\mbox{cm}^2$, which opens up two opportunities:
First, nonlinear light-matter interaction can be investigated using micrometer-thin samples~\cite{Sommer_Nature_2016}, where propagation effect play a minor role.
Second, \emph{nonperturbative} nonlinear phenomena become accessible to time-resolved measurements.
Examples of such processes include the Franz-Keldysh effect \cite{Franz_Nat_Forsch_1934_Absorptionkante,Keldysh_JETP_1958_Franz-Keldysh_Effect,Otobe_PRB_2016_Franz-Keldysh}, interband tunneling \cite{Corkum_et_all_JOPB_Attosecond_Ionization_SiO2}, and high-harmonic generation \cite{Ghimire2012PhysRevA.85.043836_HHG_in_Crystals}.

In the following, we consider pump-probe measurements where a circularly polarized few-cycle infrared (IR) pump pulse impinges at normal incidence on a uniaxial centrosymmetric crystal (sapphire)
along the optical axis.
In this geometry, the linear optical properties of the crystal are effectively isotropic.
The induced optical Faraday effect is probed by a linearly polarized ultraviolet (UV) pulse that is significantly shorter than the pump pulse, but not necessarily shorter than its optical cycle.
The probe pulse is assumed to be sufficiently weak to neglect all nonlinear processes that involve more than one UV photon.
For practical reasons, it may be beneficial to use a noncollinear geometry to spatially separate the propagated probe pulse from UV light that the pump pulse may generate without assistance from the probe pulse.
Even in this case, the angle may be chosen small enough to neglect it while modeling propagation in a micrometer-thin sample.

\section{Model}
We simulate electron dynamics in three spatial dimensions, working in the basis of stationary Bloch states in the velocity gauge.
Within the dipole approximation, equations that describe electron dynamics at different crystal momenta $\bm{k}$ are decoupled from each other:
\begin{equation}
  \iu \hbar \dfrac{\de}{\de t}\hat \rho_{\bm{k}} = \left[ \hat{H}^0_{\bm{k}} + \frac{e}{m_e}\bm{A}(t) \cdot \hat{\bm{p}}_{\bm{k}}, \hat \rho_{\bm{k}} \right]. 
  \label{eq:EOM}
\end{equation}
Here, the electric field $\bm{F}(t)$ acting on electrons enters Eq.~\eqref{eq:EOM} via $\bm{A}(t)=-\int_{-\infty}^t \bm{F}(t')\,dt'$, $e>0$ is elementary charge, and $m_e$ is electron mass.
We constructed the unperturbed Hamiltonian $H^0_{\bm{k}}$ from 36 valence bands (VB) and 160 conduction bands (CB) obtained in density-functional-theory calculations, which we performed for $\text{Al}_2\text{O}_3$ using Wien2k \cite{Schwarz_CPC_2002}.
The large number of bands is characteristic of velocity-gauge calculations \cite{Wu_PRA_2015,Yakovlev_Springer_2016}.
The momentum matrix elements, $\hat{\bm{p}}_{\bm{k}}$, were likewise obtained from Wien2k.
We used the modified Becke-Johnson exchange-correlation potential, which yields a band gap of $E_{\mathrm{g}} = 8.8$~eV, in agreement with experimental observations \cite{Olivier_Surface_1981}.

At the beginning of a simulation, $\hat \rho_{\bm{k}}$ represents a state where all valence bands are incoherently occupied, while the conduction-band states are empty.
We calculate the polarization response, $\bm{P}(t)$, by
solving Eq.~\eqref{eq:EOM} (technically, we solve an equivalent system of time-dependent Schr\"odinger equations), evaluating the electric current density, $\bm{J}(t)$, and integrating it with respect to time: $\bm{P}(t) = \int_{-\infty}^t \bm{J}(t')\,dt'$.
A particular Cartesian component, $\ell \in \{x,y,z\}$, of the current density is evaluated as
\begin{equation}
\label{eq:current}
{J}_{\ell}(t) =  - \frac{2 e}{m_e}
  \left(\frac{e \mathcal{N}_{\ell}}{ V_{\mathrm{cell}}} {A}_{\ell}(t) + \int_{\mathrm{BZ}}\! \frac{\de^3 k}{\left(2\pi \right)^3}\,  \mathrm{Tr} \left[ \hat{\rho}_{\bm{k}}(t)  \hat{{p}}_{\ell,\bm k} \right]\right),
\end{equation}
where $ V_{\mathrm{cell}}$ is the unit-cell volume, and
\begin{equation}
\mathcal{N}_{\ell} = \dfrac{2}{m_e}\sum_{i \in \mathrm{CB}} \sum_{j \in \mathrm{VB}} \frac{|p_{\ell,ij}|^2}{E_i-E_j}
\end{equation}
is the effective number of electrons per unit cell \cite{Ambrosch_CPC_2006}.
Since numerical calculations always use a truncated set of bands, the Thomas--Reiche--Kuhn sum rule is not exactly satisfied, which leads to problems such as the divergence of the linear polarization response in the low-frequency limit~\cite{Aversa_PRB_1995}.
Using the effective number of electrons in \eqref{eq:current}, we compensate for the violation of this sum rule and reduce the number of bands required for numerical convergence~\cite{Yakovlev_CPC_2017}.

We define the electric fields of the pulses via
\begin{align}
 \bm{F}^{\mathrm{P}}(t) &= \Re \left[ f^{\mathrm{P}}(t) e^{-\iu \omega_{\mathrm{P}} t } \bm{u}_{\mathrm{P}} \right] = -\de\bm{A}^{\mathrm{P}}/\de t,\\
\bm{A}^{\mathrm{P}}(t) &=
 F_{\mathrm{P}} \omega_{\mathrm{P}}^{-1}
 \ee^{-2\ln (2) t^2 / T_{\mathrm{P}}^2}
 \Re \left[ \iu e^{-\iu \omega_{\mathrm{P}} t } \bm{u}_{\mathrm{P}} \right].
\end{align}
Here, $\mathrm{P} \in \{\mathrm{IR}, \mathrm{UV}\}$, $F_{\mathrm{P}}$ is the amplitude of the electric field, $T_{\mathrm{P}}$ is the full width at half maximum of the pulse intensity, and the central pulse frequency is related to its central wavelength via $\omega_{\mathrm{P}} = 2 \pi c / \lambda_{\mathrm{P}}$, $c$ being the vacuum speed of light.
For the circularly polarized IR pump pulse, we used $\lambda_{\mathrm{IR}}=750 \: \mathrm{nm}, \: \bm{u}_{\mathrm{IR}} = (1,\iu,0)$, and $T_{\mathrm{IR}} = 5$~fs.
For the linearly polarized UV probe pulse, we used $\lambda_{\mathrm{UV}}=250 \: \mathrm{nm}, \: \bm{u}_{\mathrm{UV}} = (1,0,0)$, $T_{\mathrm{UV}} = 2.5$~fs, and we kept the peak field strength fixed at $F_{\mathrm{UV}} = 1 \: \mathrm{mV/}$\AA.
Modeling pump-probe measurements, we introduce the delay, $\tau$, via the argument of the probe field: $\bm{F}^{\mathrm{UV}}(t - \tau)$.
In the following, we use Fresnel's formula, $F_{\mathrm{P}}^{\mathrm{vac}} = \frac{1}{2} [1+n(\omega_{\mathrm{P}})] F_{\mathrm{P}}$, as an approximate relation between the vacuum amplitudes of the incident pulses, $F_{\mathrm{P}}^{\mathrm{vac}}$, and those inside the crystal.

Investigating how the polarization state of the probe pulse changes during the propagation along the $z$ axis, we define its polarization angle $\theta(\omega,z)$ and ellipticity $\epsilon(\omega,z) = \tan \alpha (\omega,z) $ in the frequency domain via 
\begin{equation}
\label{eq:polarization_state}
\frac{\left(F_x^{\mathrm{UV}}\right)^* F_y^{\mathrm{UV}}}{\left|F_x^{\mathrm{UV}}\right|^2 + \left|F_y^{\mathrm{UV}}\right|^2} =\\
\frac{\cos \bigl(2\alpha\bigr) \sin\bigl(2 \theta \bigr) + \iu \sin\bigl(2 \alpha \bigr)}{2}.
\end{equation}
If the probe pulse is initially polarized along the $x$ axis, we obtain (see Appendix~\ref{AppendixB})
\begin{equation}
\label{eq:induced_chirality}
\left(\frac{\partial \theta}{\partial z} + \iu \frac{\partial \epsilon}{\partial z}\right)\Biggr|_{z=0} =
 \frac{2 \pi \iu \omega
 	\left(F_x^{\mathrm{UV}}(\omega,0)\right)^* P_y(\omega,0)}{c n(\omega) \left| F_x^{\mathrm{UV}}(\omega,0) \right|^2},
\end{equation}
where $n(\omega)$ is the refractive index (see Appendix~\ref{AppendixB} for more details).

\section{Results and discussion}
\begin{figure}[!htb]
	\includegraphics[width=0.9\columnwidth]{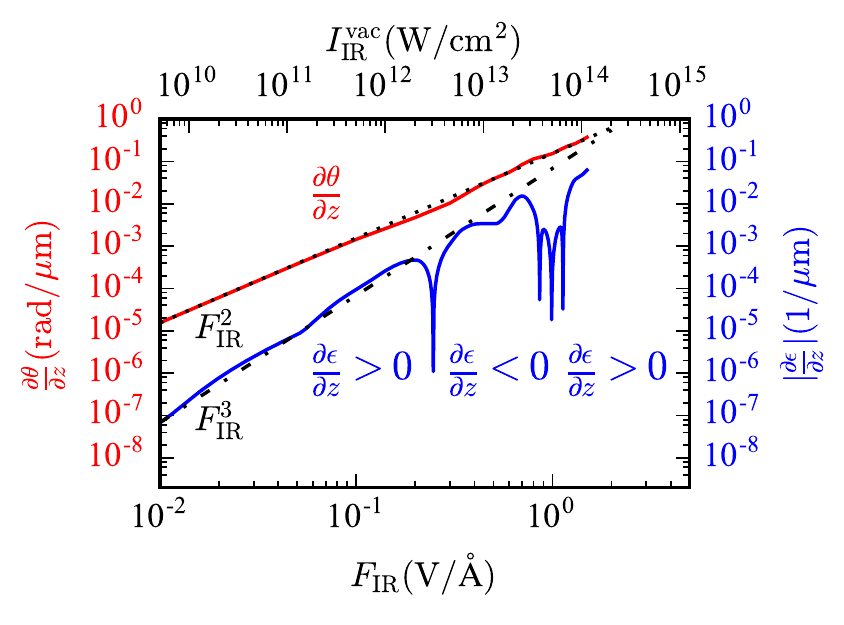}
	\caption{The induced polarization rotation and circular dichroism at the central UV frequency for the zero delay between the pulses.
		The upper horizontal axis is labeled with peak intensities of the incident IR pulse in vacuum.
		The lower horizontal axis shows the peak IR field at the crystal surface.}
	\label{fig:field_scaling}
\end{figure}

Presenting our results, we first show how the optical Faraday effect depends on the strength of the IR field.
The red curve in Fig.~\ref{fig:field_scaling} illustrates that deviations from the  $\partial \theta / \partial z \propto F_{\mathrm{IR}}^2$ scaling law are small even at IR intensities that are close to the damage threshold.
There results were obtained for pump and probe pulses arriving simultaneously ($\tau=0$).
For a peak IR intensity of $10^{13}\ \mbox{W}/\mbox{cm}^2$, the induced optical Faraday rotation at the central UV frequency is $0.03$ radians ($1.7^\circ$) per micrometer.
Note that reaching this rotation strength in the conventional Faraday effect would require a magnetic field as strong as 700 Tesla, which exceeds the strongest nondestructive magnetic fields currently available in laboratories.
Consequently, the polarization rotation is not due to light-induced magnetization (inverse Faraday effect).

In addition to polarization rotation, the UV pulse also experiences IR-induced circular dichroism, shown by the blue curve in Fig.~\ref{fig:field_scaling}.
In the weak-field limit, the induced ellipticity per unit propagation length scales as $\partial \epsilon / \partial z \propto F_{\mathrm{IR}}^3$ because at least three IR photons must be absorbed in addition to a UV photon to overcome the band gap.
At $F_{\mathrm{IR}}=0.25$~V/{\AA}, the induced zero-delay ellipticity changes its sign, which looks like a narrow downward spike on the logarithmic scale.
Altogether, $\partial \epsilon / \partial z$ changes its sign four times in Fig.~\ref{fig:field_scaling}.

To clarify the origin of the induced chirality, we first review the relevant wave mixing processes within the standard framework of nonlinear optics, where the nonlinear polarization response in the vicinity of a frequency $\omega=\omega_1+\omega_2+\omega_3$ is described by the third-order susceptibility tensor $\chi_{ijkl}^{(3)}(\omega; \omega_1, \omega_2, \omega_3)$.
As long as the polarization response is linear with respect to the probe field, the optical Faraday effect is best understood by decomposing the linearly polarized UV pulse into its circularly polarized components: 
$\bm{f}^{\mathrm{UV}}(t) = \bm{e}_{+} f_{+}^{\mathrm{UV}}(t) + \bm{e}_{-} f_{-}^{\mathrm{UV}}(t)$ with $\bm{e}_{\pm} = (\hat{\bm{x}} \pm \iu \hat{\bm{y}}) / \sqrt{2}$.
In the case  $\bm{f}^{\mathrm{UV}}(t) \parallel \hat{\bm{x}}$, we have $f_{+}^{\mathrm{UV}}(t) = f_{-}^{\mathrm{UV}}(t) = \hat{\bm{x}} \cdot \bm{f}^{\mathrm{UV}}(t) / \sqrt{2}$.
The pump pulse rotates the polarization plane of the probe pulse if it has different effects on its left- and right-rotating components.
If the two components experience different absorption, circular dichroism is observed.

\begin{figure}[!htb]
	\begin{tabular}{p{0.9\columnwidth} p{0pt}}
		\vspace{-5mm} \includegraphics[width=0.9\columnwidth]{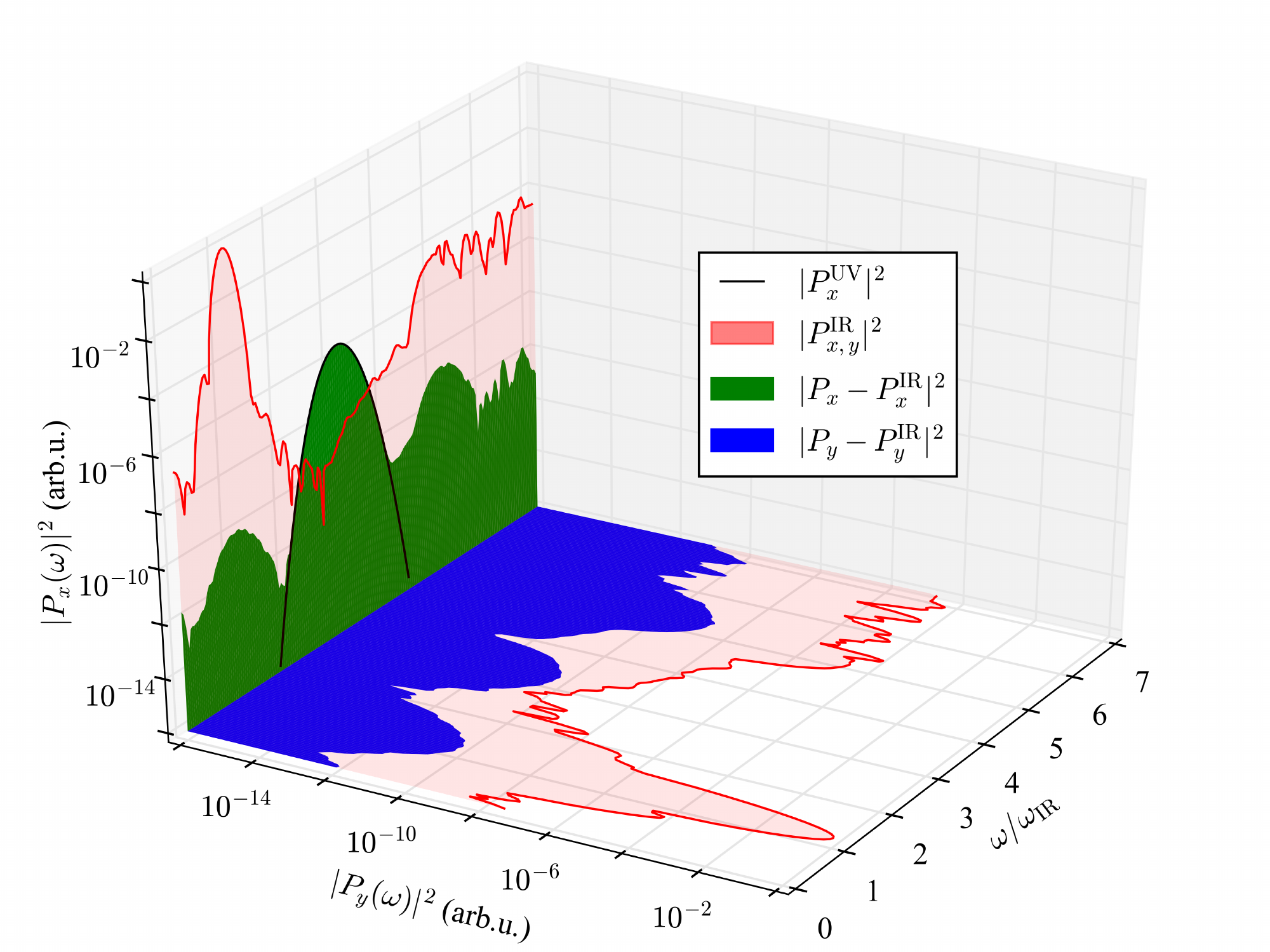} &
		\vspace{2mm} \hspace{-0.9\columnwidth}
	    \textbf{(a)}
	\end{tabular}
	\begin{tabular}{p{0.9\columnwidth} p{0pt}}
		\vspace{0pt} \includegraphics[width=0.8\columnwidth]{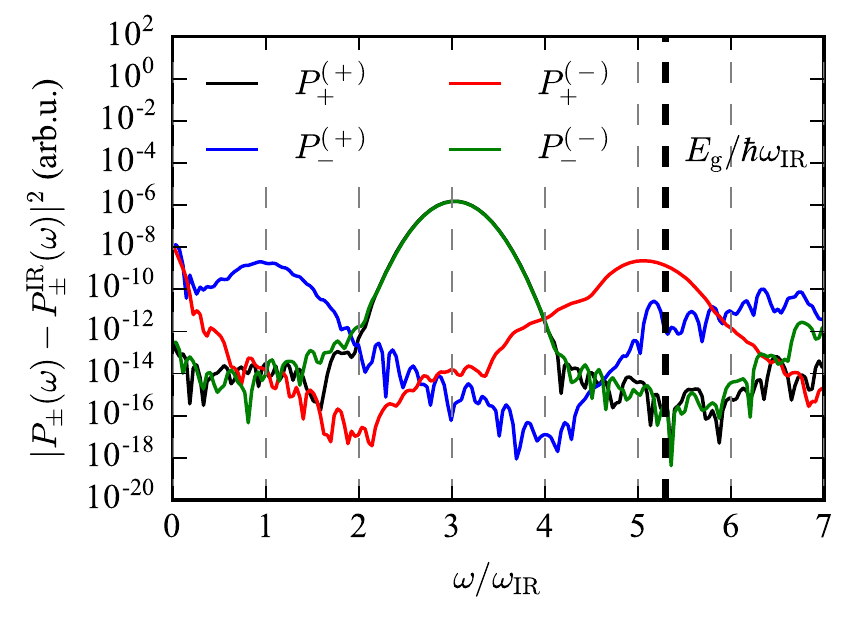} &
		\vspace{1mm} \hspace{-0.9\columnwidth}
		\textbf{(b)}
	\end{tabular}
	\caption{The spectral analysis of the polarization response for $F_{\mathrm{IR}} = 1$~V/{\AA} and the zero pump-probe delay.
	(a) The $x$ and $y$ components of the IR-only response $\bm{P}^{\mathrm{IR}}$ (pale red), the UV-only response $\bm{P}^{\mathrm{UV}}$ (black curve), and the polarization induced by both pulses after subtracting the IR-only response: $\bm{P} - \bm{P}^{\mathrm{IR}}$ (the green and blue areas).
	All the curves are plotted on the same scale.
	(b) The decomposition of $\bm{P} - \bm{P}^{\mathrm{IR}}$ into the left- and right-rotating components induced by the left- and right-rotating components of the UV pulse (see the text for further details).
	The vertical black dashed line shows the position of the band edge.}
	\label{fig:polarization_spectra}
\end{figure}

Wave mixing processes that involve one UV photon and two IR photons also produce light at frequencies $\omega_{\mathrm{UV}} \pm 2 \omega_{\mathrm{IR}}$.
Since the duration of our bandwidth-limited probe pulse is comparable to the oscillation period of the pump field, the three wave-mixing channels are spectrally separated (see Fig.~\ref{fig:polarization_spectra}).
	In this case, the polarization response to each circularly polarized component of the probe pulse, $P_{\pm}(t, \omega_{\mathrm{UV}}) = \chi_{\pm}^{\mathrm{eff}} f_{\pm}^{\mathrm{UV}}(t)$, is described by effective susceptibilities: $\chi_{\pm}^{\mathrm{eff}} = \chi^{(1)} + \Delta\chi_{\pm}$.
For an IR field rotating counter-clockwise, we obtain (see Appendix~\ref{AppendixA})
\begin{multline}
\label{eq:chi_chirality}
\Delta\chi_{\pm} = 12 F_{\mathrm{IR}}^2 \bigl[ \chi_{1 1 1 1}^{(3)}(\omega_{\mathrm{UV}}; -\omega_{\mathrm{IR}}, \omega_{\mathrm{IR}}, \omega_{\mathrm{UV}})\\
- \chi_{2 2 1 1}^{(3)}(\omega_{\mathrm{UV}}; \mp\omega_{\mathrm{IR}}, \pm\omega_{\mathrm{IR}}, \omega_{\mathrm{UV}})\bigr]
\end{multline}
(replace $\Delta\chi_{\pm}$ with  $\Delta\chi_{\mp}$ for a clockwise-rotating IR field).
Optical Faraday effect emerges if $\Delta\chi_{-} \ne \Delta\chi_{+}$, which requires
$\chi_{2 2 1 1}^{(3)}(\omega_{\mathrm{UV}}; \omega_{\mathrm{IR}}, -\omega_{\mathrm{IR}}, \omega_{\mathrm{UV}}) \ne \chi_{2 2 1 1}^{(3)}(\omega_{\mathrm{UV}}; -\omega_{\mathrm{IR}}, \omega_{\mathrm{IR}}, \omega_{\mathrm{UV}})$.
The two susceptibilities are equal if the polarization response is instantaneous (more precisely, if Kleinman's symmetry holds).
Therefore, the optical Faraday effect is a consequence of the nonlinear polarization response being non-instantaneous.

We now turn our attention to the case of a field as strong as $F_{\mathrm{IR}} = 1$~V/{\AA}, where third-order susceptibilities no longer provide an accurate description of the optical Faraday effect (see Fig.~\ref{fig:field_scaling}).
Fig.~\ref{fig:polarization_spectra} shows that, even in this case, the three wave-mixing channels mentioned above are well separated from each other.
In Fig.~\ref{fig:polarization_spectra}(a), we use the logarithmic scale for polarization spectra, plotting the $x$- and $y$-components along the vertical and horizontal axes, respectively.
The black curve represents the polarization induced by a sole UV pulse (note that we chose $\omega_{\mathrm{UV}} = 3 \omega_{\mathrm{IR}}$).
The area filled with the pale red color represents the polarization induced by the IR pulse alone.
It illustrates that a circularly polarized IR pulse propagating along the optical axis of sapphire generates no third harmonic even if the pulse is strong enough to excite some electrons from valence into conduction bands.
The presence of such excitations is evident from the polarization response at frequencies above the band edge ($\omega / \omega_{\mathrm{IR}} \gtrsim 5.3$ in our simulations) \cite{Yakovlev_NJP_2013_Quantum_Beats_in_Dielectrics}.
Nevertheless, $|\bm{P}^{\mathrm{IR}}(\omega)|^2$ reaches significant values at frequencies close to $3 \omega_{\mathrm{IR}}$, which is why suppressing the IR-only response by using a noncollinear geometry may be required in experiments unless this response is sufficiently suppressed by phase matching (the optical Faraday effect is self-phase-matched).
We model this suppression by subtracting the IR-only response.
The areas filled with green and blue colors show $|P_x(\omega)-P_x^{\mathrm{IR}}(\omega)|$ and $|P_y(\omega)-P_y^{\mathrm{IR}}(\omega)|$, respectively.

To verify that the $\omega_{\mathrm{UV}} \pm 2 \omega_{\mathrm{IR}}$ channels have no significant effect on the polarization response at $\omega = \omega_{\mathrm{UV}}$, we decompose both the UV pulse and the polarization response into components with positive and negative helicities: $\bm{P}(\omega)-\bm{P}^{\mathrm{IR}}(\omega) = \bm{e}_{+} P_{+}(\omega) + \bm{e}_{-} P_{-}(\omega)$.
The result is shown in Fig.~\ref{fig:polarization_spectra}(b).
The subscript in $P_{\pm}^{(\pm)}$ refers to the helicity of the polarization response, while the superscript denotes the helicity of the probe pulse.
Analyzing these components we see that, well below the band edge, they are consistent with the conservation of the spin angular momentum of a photon, $\bm{S}$.
For example, $P_{-}^{(+)}$, which is the blue line, shows the clockwise rotating component of the polarization response that would be generated by a counter-clockwise rotating UV field.
This component peaks at $\omega_{\mathrm{UV}} - 2 \omega_{\mathrm{IR}} = \omega_{\mathrm{IR}}$ because a parametric third-order process where one UV photon with $S_z=1$ is absorbed and two IR photons with $S_z=-1$ are emitted must generate a photon with $S_z=-1$.
However, these considerations would prohibit the blue curve from peaking at $\omega_{\mathrm{UV}} = 2 \omega_{\mathrm{IR}} = 5 \omega_{\mathrm{IR}}$ due to $|S_z| \le 0$.
Near the band edge, the conservation of the photon spin is violated because some angular momentum is transferred to charge carriers.

\begin{figure}[!htb]
	\begin{tabular}{p{0.9\columnwidth} p{0pt}}
		\vspace{2mm} \includegraphics[width=0.9\columnwidth]{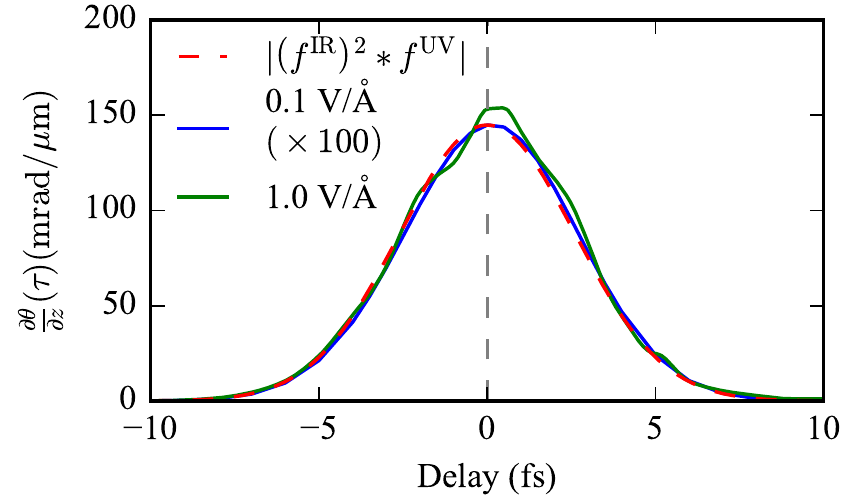} &
		\vspace{0pt} \hspace{-0.9\columnwidth}
		\textbf{(a)}
	\end{tabular}\\[-5mm]
	\begin{tabular}{p{0.9\columnwidth} p{0pt}}
		\vspace{2mm} \includegraphics[width=0.9\columnwidth]{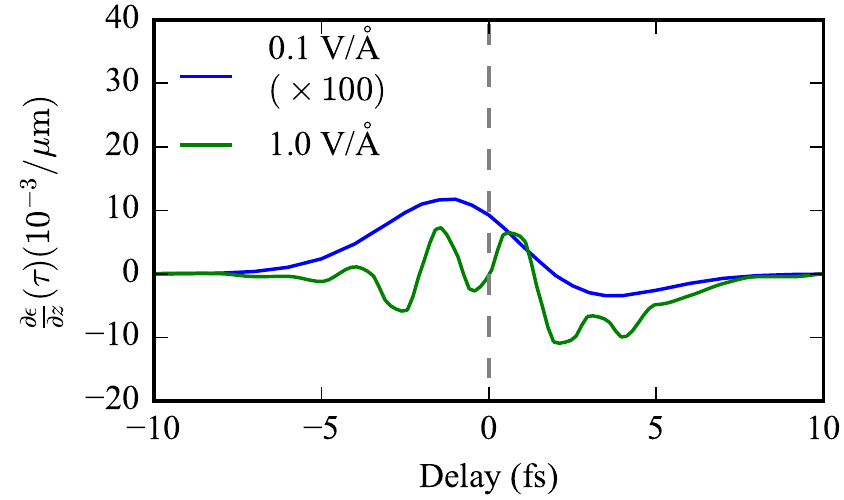} &
		\vspace{0pt} \hspace{-0.9\columnwidth}
		\textbf{(b)}
	\end{tabular}
	\caption{The dependence of induced (a) polarization rotation and (b) ellipticity of the probe pulse on the pump-probe delay.
	For a positive delay, the probe pulse arrives after the pump pulse.
	In both panels, data for $F_{\mathrm{IR}}=0.1$~V/{\AA} (blue curves) was multiplied by a factor of 100 to make it comparable to that for $F_{\mathrm{IR}}=1$~V/{\AA} (green curves).
	In panel (a), the dashed red curve shows the convolution between the UV envelope and the square of the IR envelope.}
	\label{fig:delay_scan}
\end{figure}
Since the optical Faraday effect results from the nonlinear medium response being non-instantaneous, these effects lend themselves to pump-probe measurements.
In Fig.~\ref{fig:delay_scan}(a), we show the delay dependence of the polarization rotation evaluated at the central UV frequency.
At $F_{\mathrm{IR}} = 0.1$~V/{\AA}, the conventional perturbative nonlinear optics works well (see Fig.~\ref{fig:field_scaling}), so we expect $\partial \theta / \partial z \propto F_{\mathrm{IR}}^2$.
It is therefore not surprising that $\partial \theta / \partial z$ as a function of the delay (blue curve) has precisely the same shape as the convolution $\left|\left(f^{\mathrm{IR}}\right)^2 * f^{\mathrm{UV}}\right|$ (red dashed curve).
When we increase $F_{\mathrm{IR}}$ to 1~V/{\AA}, we observe a small but significant reshaping of $\partial \theta / \partial z$, revealing the onset of nonadiabatic processes that disappear by the end of the pulse.
Our analysis of the results presented in Fig.~\ref{fig:polarization_spectra} suggests that these effects are probably related to the creation of real (nonvirtual) electronic excitations and angular momentum transfer from light to charge carriers.
For 1~V/{\AA}, the residual excitation density is $2.9 \times 10^{-5}$ electrons per unit cell.
We also note that the dynamical Franz-Keldysh effect becomes an important excitation mechanism at such field strengths; nonadiabatic features of this effect have recently been predicted in numerical simulations \cite{Otobe_PRB_2016_Franz-Keldysh}, and they are likely to contribute to the induced circular dichroism~\cite{Citrin_PRB_1999}.
We point out, however, that the Franz-Keldysh effect is not a prerequisite for observing induced circular dichroism, as evident from the $\propto F_{\mathrm{IR}}^3$ scaling in the weak-field limit (see Fig.~\ref{fig:field_scaling}).

The nonadiabatic effects manifest themselves more vividly in the delay dependence of the induced ellipticity of the UV pulse.
In Fig.~\ref{fig:delay_scan}(b), we show $\partial \epsilon / \partial z$ evaluated with the aid of Eq.~\eqref{eq:induced_chirality}.
When the IR field is weak (0.1~V/{\AA}), the induced ellipticity is mainly due to the time-dependent polarization rotation---the intensity of the 5-fs IR pulse changes significantly during the 2.5-fs UV pulse, and so does $\partial \theta / \partial z$.
This contribution to the induced ellipticity represents transfer of angular momentum from the pump pulse to the probe one in the absence of circular dichroism, and it is particularly large at delays where $\de |f^{\mathrm{IR}}(\tau)|^2/ \de \tau$ is large.
At the same time, even this relatively weak IR field induces some noticeable ellipticity at $\tau=0$, where $\de |f^{\mathrm{IR}}(\tau)|^2/ \de \tau = 0$, and, therefore, the main mechanism is the induced circular dichroism (helicity-dependent absorption).
This contribution has a qualitatively different delay dependence: it does not change sign at $\tau=0$.
The sum of the two delay-dependent functions, one of which is approximately odd and the other one is approximately even, results in the asymmetric shape of the blue curve. 
At $F_{\mathrm{IR}} = 1$~V/{\AA} (green curve), the induced ellipticity exhibits an oscillatory dependence on the pump-probe delay. 
The period of these oscillations is close to the optical period of the IR field (2.5~fs).
Therefore, the oscillations cannot be due to the interference with the $\omega_{\mathrm{UV}} \pm 2 \omega_{\mathrm{IR}}$ channels, which would result in a period of $\sim 2\pi/(2 \omega_{\mathrm{IR}}) \approx 1.3$~fs.
These oscillations represent a strong-field effect, and their phase depends on $F_{\mathrm{IR}}$.

\section{Conclusions}
We conclude this paper with a brief discussion of the role played by physical symmetries.
We have considered the case where the polarization response to the probe pulse alone (without the strong field) is invariant with respect to parity and time-reversal symmetries.
A strong adiabatic (i.e., with frequency within the band gap) circularly-polarized optical field transiently relaxes both these symmetries.
In analogy to the conventional Faraday effect, imposing time reversal on the crystal and the applied (strong) field reverses the polarization rotation of the probe pulse.
However, in contrast to the conventional Faraday effect, the time reversal symmetry is relaxed not due to the presence of magnetic field but due to a transient transfer of angular momentum from light to matter.
Even though the optical Faraday effect is a consequence of the nonlinear polarization response being non-instantaneous, the response time appears to be so small that the optical Faraday effect in solids is inertialess for any practical purposes.
Indeed, the time span of the pump-probe delay dependence in Fig.\ \ref{fig:delay_scan} is the same as the duration of the pump pulse even if, within the pulse, the excitation of electrons to conduction bands changes the shapes of $\theta(\tau)$ and $\epsilon(\tau)$.
Possibly, averaging over crystal momenta in Eq.~\eqref{eq:current} leads to effective collisionless dephasing (Landau damping) that counteracts excitation-induced chirality.
We have studied the ultrafast optical Faraday effect for a uniaxial crystal, but our general conclusions are also valid for isotropic media.

For basic research, the ultrafast optical Faraday effect is attractive as a spectroscopic tool capable of studying chiral dynamics with an attosecond temporal resolution.
Potential applications of this effect include ultrafast all-optical  circular-polarization modulators, optical isolators, and optical circulators without a need for magnetic field.

\begin{acknowledgments}
M.\,S.\,W.\ was supported by the International Max Planck Research School of Advanced Photon Science (IMPRS-APS).
V.\,S.\,Y.\ was supported by the DFG Cluster of Excellence: Munich-Centre for Advanced Photonics.
His work at the GSU and CeNO on perturbative theory and preliminary numerical modeling was supported by a grant DE-SC0007043 from the Physical Behavior of Materials Program, Office of Basic Energy Sciences, U.S. Department of Energy.
M.\,I.\,S.\ was supported by a MURI grant FA9550-15-1-0037 from the U.S. Airforce Office of Scientific Research.
The authors gratefully acknowledge useful discussions with S.~Kruchinin.
\end{acknowledgments}

\appendix

\section{Third-order polarization response}
\label{AppendixA}
In this section, we provide some expressions for the third-order nonlinear polarization induced by a circularly polarized IR pulse and a weak linearly polarized UV pulse.
We obtained these equations using the standard methods of nonlinear optics implemented in a Mathematica script.
Doing so, we used the spatial symmetries of the $\chi_{ijkl}^{(3)}$ tensor that correspond to a crystal with the sapphire symmetry ($\bar{3}2/\mbox{m}$).
We assume that both laser beams propagate along the crystal axis, which we chose to be the $z$ axis of our coordinate system.

In the main text, we defined the complex pulse amplitude via
$\bm{F}^{\mathrm{P}}(t) = \Re \left[ f^{\mathrm{P}}(t) e^{-\iu \omega_{\mathrm{P}} t } \bm{u}_{\mathrm{P}} \right]$, where $\mathrm{P} \in \{\mathrm{IR}, \mathrm{UV}\}$.
Here, to be consistent with the notation most frequently used in nonlinear optics, we use a different definition:
\begin{equation}
\bm{F}^{\mathrm{P}}(t) = f^{\mathrm{P}}(t) e^{-\iu \omega_{\mathrm{P}} t } \bm{u}_{\mathrm{P}} + \text{c.c.}
\end{equation}
We used an IR pulse with the positive helicity: $\bm{u}_{\mathrm{IR}} = (1,\iu,0)$.
Even though we propose measurements with a linearly polarized probe pulse, it is instructive to decompose the pulse into its left- and right-rotating circularly polarized components: $\bm{u}_{\mathrm{IR}} = (1,\pm\iu,0)$.
The sign on the right-hand side of this expression appears in the subscript of $P_{\pm}^{(\pm)}(t)$ in the equations below, where we use the same convention as in the main text: the superscript refers to the helicity of the probe pulse, while the subscript refers to the helicity of the polarization response.
At the central frequency of the UV pulse, we obtained the following expressions for the part of the third-order polarization response that mixes the IR and UV beams:
\begin{multline}
  P_{+}^{(+)}(t; \omega_{\mathrm{UV}}) = 12 \sqrt{2} \left[f^{\mathrm{IR}}(t)\right]^2 f^{\mathrm{UV}}(t) \\
  \times \Bigl[
    \chi_{1 1 1 1}^{(3)} (\omega_{\mathrm{UV}}; -\omega_{\mathrm{IR}}, \omega_{\mathrm{IR}}, \omega_{\mathrm{UV}}) \\-
    \chi_{2 2 1 1}^{(3)} (\omega_{\mathrm{UV}}; -\omega_{\mathrm{IR}}, \omega_{\mathrm{IR}}, \omega_{\mathrm{UV}})
  \Bigr] + \text{c.c.},
\end{multline}
\begin{multline}
  P_{-}^{(-)}(t; \omega_{\mathrm{UV}}) = 12 \sqrt{2} \left[f^{\mathrm{IR}}(t)\right]^2 f^{\mathrm{UV}}(t) \\
  \times \Bigl[
  \chi_{1 1 1 1}^{(3)} (\omega_{\mathrm{UV}}; -\omega_{\mathrm{IR}}, \omega_{\mathrm{IR}}, \omega_{\mathrm{UV}}) \\-
  \chi_{2 2 1 1}^{(3)} (\omega_{\mathrm{UV}}; \omega_{\mathrm{IR}}, -\omega_{\mathrm{IR}}, \omega_{\mathrm{UV}})
  \Bigr] + \text{c.c.},
\end{multline}
\begin{equation}
  P_{+}^{(-)}(t; \omega_{\mathrm{UV}}) = P_{-}^{(+)}(t; \omega_{\mathrm{UV}}) = 0.
\end{equation}
Deriving these equations, we dropped terms that were nonlinear with respect to $f^{\mathrm{UV}}(t)$ because the UV pulse is assumed to be weak.
Deviating from the notation used in the main text, we do not explicitly account for the delay.
If the UV pulse is delayed by $\tau$, its envelope $f^{\mathrm{UV}}(t)$ must be replaced with $f^{\mathrm{UV}}(t - \tau) \ee^{\iu \omega_{\mathrm{UV}} \tau}$.

Absorbing a UV photon and two IR photons generates the following components of the nonlinear polarization:
\begin{multline}
P_{+}^{(-)}(t; \omega_{\mathrm{UV}} + 2 \omega_{\mathrm{IR}}) = 12 \sqrt{2} \left[f^{\mathrm{IR}}(t)\right]^2 f^{\mathrm{UV}}(t) \\
\times \chi_{2 2 1 1}^{(3)} (\omega_{\mathrm{UV}} + 2 \omega_{\mathrm{IR}}; \omega_{\mathrm{IR}}, \omega_{\mathrm{IR}}, \omega_{\mathrm{UV}})
 + \text{c.c.},
\end{multline}
\begin{multline}
  P_{+}^{(+)}(t; \omega_{\mathrm{UV}} + 2 \omega_{\mathrm{IR}}) = 
  P_{-}^{(-)}(t; \omega_{\mathrm{UV}} + 2 \omega_{\mathrm{IR}}) \\= 
  P_{-}^{(+)}(t; \omega_{\mathrm{UV}} + 2 \omega_{\mathrm{IR}}) = 0.
\end{multline}

Absorbing a UV photon and emitting two IR photons generates
\begin{multline}
P_{-}^{(+)}(t; \omega_{\mathrm{UV}} - 2 \omega_{\mathrm{IR}}) = 12 \sqrt{2} \left[f^{\mathrm{IR}}(t)\right]^2 f^{\mathrm{UV}}(t) \\
\times \chi_{2 2 1 1}^{(3)} (\omega_{\mathrm{UV}} - 2 \omega_{\mathrm{IR}}; -\omega_{\mathrm{IR}}, -\omega_{\mathrm{IR}}, \omega_{\mathrm{UV}})
+ \text{c.c.}
\end{multline}
and
\begin{multline}
P_{+}^{(+)}(t; \omega_{\mathrm{UV}} + 2 \omega_{\mathrm{IR}}) = 
P_{-}^{(-)}(t; \omega_{\mathrm{UV}} + 2 \omega_{\mathrm{IR}}) \\= 
P_{+}^{(-)}(t; \omega_{\mathrm{UV}} + 2 \omega_{\mathrm{IR}}) = 0.
\end{multline}

The $x$- and $y$-components of the polarization response can be evaluated as
\begin{align}
  P_{x} &= \left( P_{+} + P_{-} \right) / \sqrt{2},\\
  P_{y} &= \iu \left( P_{+} - P_{-} \right) / \sqrt{2}.
\end{align}
Using our Mathematica script, we explicitly verified that there is no third-harmonic generation by the circularly polarized IR pulse, even if we take the fifth-order terms into account.

\subsection{The conservation of the angular momentum}
In a homogeneous isotropic medium, the spin angular momentum of a photon, $\bm{S}$, is conserved.
From this principle, the following selection rules for third-order processes follow:
(i) a circularly polarized IR field cannot generate the third harmonic;
(ii) a circularly polarized component of the UV pulse induces a circularly polarized $\bm{P}^{(3)}(\omega_{\mathrm{UV}})$ rotating in the same direction as the UV field (in other words, absorbing and emitting an IR photon as a part of wave mixing does not change the spin angular momentum of the UV light);
(iii) in the case of co-rotating IR and UV fields, emission at $\omega_{\mathrm{UV}} + 2 \omega_{\mathrm{IR}}$ is forbidden due to $S \le 1$, while $\bm{P}^{(3)}(\omega_{\mathrm{UV}} - 2 \omega_{\mathrm{IR}})$ rotates in the direction opposite to that of the light fields;
(iv) in the case of counter-rotating IR and UV fields, emission at $\omega_{\mathrm{UV}} - 2 \omega_{\mathrm{IR}}$ is forbidden, while $\bm{P}^{(3)}(\omega_{\mathrm{UV}} + 2 \omega_{\mathrm{IR}})$ rotates in the same direction as the IR field.
We have verified that the same rules apply to the $\bar{3}2/\mbox{m}$ crystal system of sapphire if the laser beam is aligned with its threefold rotation-inversion axis. This is a nontrivial fact: for example, third-harmonic generation with circularly polarized light is allowed in cubic crystals, where the linear response is also isotropic.

\section{Propagation model}
\label{AppendixB}
To evaluate the induced ellipticity and polarization rotation of the probe pulse, we need to model its propagation.
For this purpose, we employed  the first-order propagation equation in the slowly-evolving wave approximation~\cite{Brabec_RMP_2000}:
\begin{equation}
\label{eq:FOP}
  \frac{\partial \bm{F}}{\partial z} = \iu k(\omega) \bm{F}(z, \omega) +
  \frac{2 \pi \iu \omega}{c n(\omega)} \bm{P}^{\mathrm{NL}}(z,\omega).
\end{equation}
Here, we use CGS units, neglect diffraction, define the Fourier transform according to
\begin{equation}
\mathcal{F}[f(t)] = \int_{-\infty}^{\infty} f(t) \ee^{\iu \omega t}\,\de t,
\end{equation}
and define the nonlinear polarization via
\begin{equation}
\bm{P}(z, \omega) = \hat{\chi}^{(1)}(\omega) \bm{F}(z, \omega) +
 \bm{P}^{\mathrm{NL}}(z, \omega).
\end{equation}
The wave vector for propagation along the crystal axis is given by
\begin{equation}
k(\omega) = \frac{\omega}{c} n(\omega),
\end{equation}
where $n(\omega) = \sqrt{1 + 4 \pi \chi^{(1)}(\omega)}$ is the refractive index.

To obtain Eq.~\eqref{eq:induced_chirality} in the main text, we first translate Eq.~\eqref{eq:polarization_state} into
\begin{align}
\label{eq:helicity}
\alpha(\omega,z) &= \frac{1}{2} \mbox{arcsin} \left( \frac{2 \Im \left[ \left(F_x^{\mathrm{UV}}(\omega,z)\right)^* F_y^{\mathrm{UV}}(\omega,z) \right]}{|\bm{F}^{\mathrm{UV}}(\omega,z)|^2} \right),\\
\label{eq:polarization_angle}
\theta(\omega,z) &= \frac{1}{2} \mbox{arcsin} \left( \frac{2 \Re \left[ \left(F_x^{\mathrm{UV}}(\omega,z)\right)^* F_y^{\mathrm{UV}}(\omega,z) \right]}{|\bm{F}^{\mathrm{UV}}(\omega,z)|^2 \cos\bigl( 2 \alpha(\omega,z) \bigr)} \right),
\end{align}
and
\begin{equation}
\epsilon(\omega,z) = \tan\bigl( \alpha(\omega,z) \bigr).
\end{equation}
We then consider a UV pulse that is initially polarized along the $x$-axis and notice that
\begin{equation}
\frac{\partial}{\partial z} \left(\frac{ \left( F_x^{\mathrm{UV}} \right)^{\ast} F_y^{\mathrm{UV}}}{\left| \bm F^{\mathrm{UV}} \right|^2} \right) \Biggr|_{z=0} =
\left(\frac{\partial \theta}{\partial z} + \iu \frac{\partial \alpha}{\partial z} \right) \Biggr|_{z=0}.
\end{equation}
With 
\begin{align}
F_y^{\mathrm{UV}}(0,\omega) &\equiv 0,\\
\frac{\partial F_x^{\mathrm{UV}}}{\partial z} \biggr|_{z=0} &= 
\iu k(\omega) F_x^{\mathrm{UV}} (0, \omega), \\
\frac{\partial F_y^{\mathrm{UV}}}{\partial z} \biggr|_{z=0} &= 
\frac{2 \pi \iu \omega}{c n(\omega)} P_y^{\mathrm{NL}}(0,\omega),\\
\frac{\partial \alpha}{\partial z}\Biggr|_{z=0} &= \frac{\partial \epsilon}{\partial z}\Biggr|_{z=0},
\end{align}
and,
\begin{multline}
\frac{\partial  \left| \bm F^{\mathrm{UV}} \right|^2}{\partial z}  \biggr|_{z=0} =
2 \text{Re} \Biggl[
\left( F_x^{\mathrm{UV}} \right)^* \frac{\partial  F_x^{\mathrm{UV}}}{\partial z} +\\
\left( F_y^{\mathrm{UV}} \right)^* \frac{\partial  F_y^{\mathrm{UV}}}{\partial z}
\Biggr] \Biggr|_{z=0} =
-\frac{2 \omega}{c}  \left| F_x^{\mathrm{UV}} (0, \omega) \right|^2 \text{Im} [ n(\omega) ],
\end{multline}
we obtain, neglecting the linear absorption ($\text{Im} [ n(\omega) ]= 0$),
\begin{equation}
\label{eq:polarization_rotation1}
\left(\frac{\partial \theta}{\partial z} + \iu \frac{\partial \epsilon}{\partial z}\right)\Biggr|_{z=0} =
\frac{2 \pi \iu \omega
  \left(F_x^{\mathrm{UV}}(\omega,0)\right)^* P_y^{\mathrm{NL}}(\omega,0)}{c n(\omega) \left| F_x^{\mathrm{UV}}(\omega,0) \right|^2}.
\end{equation}

If effective susceptibilities provide a good approximation for the nonlinear polarization at the UV frequency,  $P_{\pm}(z, \omega_{\mathrm{UV}}) = \Delta\chi_{\pm} F_{\pm}^{\mathrm{UV}}(z, \omega_{\mathrm{UV}})$, it is possible to obtain the following expression for the polarization rotation in an isotropic medium:
\begin{equation}
\label{eq:polarization_rotation2}
\frac{\de \theta(z, \omega_\mathrm{UV}) }{\de z} =
\frac{2 \pi \omega_{\mathrm{UV}}}{c} \text{Re} \left[ \frac{\Delta\chi_{-} - \Delta\chi_{+}}{n(\omega_{\mathrm{UV}})} \right].
\end{equation}
The right-hand side of this equation does not depend on the probe pulse.
Using the explicit expressions for $\Delta\chi_{\pm}$, we see that optical Faraday effect exists if $\chi_{2 2 1 1}^{(3)}(\omega_{\mathrm{UV}}; \omega_{\mathrm{IR}}, -\omega_{\mathrm{IR}}, \omega_{\mathrm{UV}}) \ne \chi_{2 2 1 1}^{(3)}(\omega_{\mathrm{UV}}; -\omega_{\mathrm{IR}}, \omega_{\mathrm{IR}}, \omega_{\mathrm{UV}})$.\\

\section{Numerical simulations}
\label{AppendixC}
We obtained the lattice constants for Al$_2$O$_3$  from \cite{Wang_JACE_1994}.
The  density-functional-theory and dynamical calculations were performed on an unshifted Monkhorst-Pack grid with $5 \times 5 \times 5$  $\bm{k}$-points.
For each $\bm{k}$-point, the initial mixed state can be written as a sum of independent valence-band wave functions:
\begin{equation}
\rho_{\bm{k}} (t)= \sum_i^{N_v} \vert \psi_{i,{\bm{k}}}(t) \rangle \langle \psi_{i,{\bm{k}}}(t) \vert.
\end{equation}

While the density-matrix description allow us to write the key equations in a compact and general form, we obtain the same results by solving the time-dependent Schr\"odinger equation (TDSE), which requires less computation.
We work in the interaction picture:
\begin{equation}
\label{eq:TDSE}
\iu \hbar \frac{d}{dt} \vert \tilde{\psi}_{i,{\bm{k}}} \rangle = \frac{e}{m_e}\bm{A}(t) \cdot \tilde{\hat{\bm{p}}}_{\bm{k}} \vert \tilde{\psi}_{i,{\bm{k}}} \rangle,
\end{equation}
where $\vert \tilde{\psi}_{i,{\bm{k}}} \rangle = e^{\iu \hat{H}_0 t/\hbar} \vert \psi_{i,{\bm{k}}} \rangle$, $ \tilde{\hat{\bm{p}}}_{\bm{k}}  = e^{\iu \hat{H}_0 t/\hbar} \hat{\bm{p}}_{\bm{k}} e^{-\iu \hat{H}_0 t/\hbar}$, and $\hat{H}_0$ is the unperturbed Hamiltonian.
We used the 4th-order Runge--Kutta scheme to solve Eq.~\eqref{eq:TDSE} for 36 valence bands and 160 conduction bands.
Thus, we had $36 \times 5 \times 5 \times 5 = 4500$ independent differential equations, each of which was solved in a basis of $36 + 160 = 196$ stationary states.
On a desktop computer (Intel Core 2 Duo E8400 3.00 GHz),  solving the TDSE for a single $\bm{k}$-point and a particular initial (valence) band takes 12 seconds.

Our simulations used the following expression for the Hamiltonian: $\hat{H}(t) = \hat{H}^0_{\bm{k}} + \frac{e}{m_e}\bm{A}(t) \cdot \hat{\bm{p}}_{\bm{k}}$.
A trivial unitary transformation relates this Hamiltonian to another velocity-gauge Hamiltonian that is frequently encountered in the literature: $\hat{H}^0_{\bm{k}} + \frac{e}{m_e}\bm{A}(t) \cdot \hat{\bm{p}}_{\bm{k}} + \frac{e^2}{m_e} A^2(t)$.
This transformation reads
\begin{equation}
|\psi\rangle = \exp\left[\frac{\iu e^2}{2 \hbar m_e} \int_{-\infty}^{t} A^2(t')\,\de t' \right] |\psi'\rangle.
\end{equation}
Therefore, dropping the $e^2 m_e^{-1} A^2(t)$ term in the Hamiltonian does not introduce an additional approximation.


%

\end{document}